\documentclass[preprint,number,sort&compress,twocolumn,3p]{elsarticle}

\pdfoutput=1
\usepackage{amssymb}    % Mathematical symbols
\usepackage{amsmath}    % More options for mathematics
\usepackage{epstopdf}   % Convert eps to pdf
\usepackage[separate-uncertainty=true]{siunitx}   % Proper formatting of units in math mode
\usepackage{color}      % Supports text color if needed
\usepackage{lmodern}    % Loading fonts
\usepackage[pdftex,hyperfootnotes=false,pdfpagelabels]{hyperref}
\usepackage{graphicx}
\usepackage{booktabs}
\usepackage{bigints}
\usepackage{slashed}
\usepackage[normalem]{ulem}
\usepackage{rotating}
\usepackage{array}

\usepackage{widetext}

\begin{document}

\title{On dark atoms, massive dark photons and millicharged sub-components}

\tnotetext[t1]{This article is registered under preprint number: TTK-20-02}

\author[add1]{Felix Kahlhoefer}
\ead{kahlhoefer@physik.rwth-aachen.de}

\author[add1,add2]{Einar Urdshals}
\ead{urdshals@physik.rwth-aachen.de}

\address[add1]{Institute for Theoretical Particle Physics and Cosmology (TTK), RWTH Aachen University, D-52056 Aachen, Germany}
\address[add2]{Institutt for fysikk,  NTNU,  Trondheim,  Norway}

\begin{abstract}

We present a simple model of two dark matter species with opposite millicharge that can form electrically neutral bound states via the exchange of a massive dark photon. If bound state formation is suppressed at low temperatures, a sub-dominant fraction of millicharged particles remains at late times, which can give rise to interesting features in the $21\,\mathrm{cm}$ absorption profile at cosmic dawn. The dominant neutral component, on the other hand, can have dipole interactions with ordinary matter, leading to non-standard signals in direct detection experiments. We identify the parameter regions predicting a percent-level ionisation fraction and study constraints from laboratory searches for dark matter scattering and dark photon decays.

\end{abstract}

\maketitle

\section{Introduction}

The defining property of dark matter (DM) is that it does not participate in electromagnetic interactions in the same way as visible matter. Nevertheless, experimental and observational constraints are consistent with the possibility that DM particles carry a tiny electromagnetic charge $\epsilon e$ with $\epsilon \ll 1$ (a so-called \emph{millicharge}) or a small electric or magnetic dipole moment~\cite{McDermott:2010pa,Weiner:2012cb,Kavanagh:2018xeh,Chu:2018qrm}. Moreover, if DM consists of several different components, the constraints on any sub-dominant component are substantially relaxed and allow for sizable electromagnetic interactions.

The idea of a sub-component of millicharged DM has recently received substantial attention because such a species would have interactions with baryons that become stronger as the Universe cools down. This makes it possible to satisfy constraints on DM-baryon interactions from the Cosmic Microwave Background (CMB), while allowing for observable effects at later times. In particular, such interactions are predicted to reduce the temperature of baryons before the beginning of reionization, leading to modifications of the $21\,\mathrm{cm}$ signal from neutral hydrogen at redshift 15--20~\cite{Munoz:2018pzp,Berlin:2018sjs,Fraser:2018acy,Barkana:2018cct,Kovetz:2018zan,Liu:2019knx}. Indeed, the EDGES experiment has recently claimed evidence for a discrepancy between predictions and observations of the $21\,\mathrm{cm}$ absorption profile at cosmic dawn~\cite{Bowman:2018yin}, and future radio telescopes will provide a wealth of data from this previously unobserved era.

In the present work we do not attempt to explain the EDGES signal, but instead study the possible origin of a subdominant component of millicharged DM. While it is often assumed that the uncharged dominant component and the millicharged sub-component are unrelated, we consider the possibility that they share a common origin and correspond to the neutral and ionized fraction of the same particle species. More specifically, we consider a dark sector comprising two types of particles with equal but opposite millicharge, called \emph{dark proton} $\mathbf{p}$ and \emph{dark electron} $\mathbf{e}$ , which can recombine to form \emph{dark hydrogen} $\mathbf{H}$.\footnote{Here and in the following we use boldface letters to denote states in the dark sector.}

The cross section for the formation of dark hydrogen via photon exchange, $\mathbf{p} + \mathbf{e} \to \mathbf{H} + A$ is proportional to $\epsilon^4 \alpha^2$, where $\alpha$ is the electromagnetic fine-structure constant, making it impossible to achieve efficient recombination from this interaction alone. We therefore assume that dark protons and dark electrons interact with each other also through the exchange of a dark photon $\mathbf{A}$ with fine structure constant $\alpha_D$, such that the bound-state formation cross section is proportional to $\alpha_D^2$. 

The cosmology of this so-called \emph{atomic DM} has been studied in detail in the literature~\cite{Kaplan:2009de,Kaplan:2011yj,CyrRacine:2012fz}. If the dark photon is massless, it is necessary for the dark sector to be substantially colder than the visible sector in order to satisfy constraints on the number of relativistic degrees of freedom during Big Bang Nucleosynthesis (BBN). Such a temperature difference is however difficult to maintain in the presence of millicharges, which are expected to lead to thermalization between the visible and the dark sector.

We therefore consider the alternative possibility that the dark photon has a mass $m_\mathbf{A} \gtrsim 10\,\mathrm{MeV}$, generated together with the DM millicharges via the Stueckelberg mechanism~\cite{Stueckelberg:1900zz}. The cross sections for bound state formation via massive dark photon exchange have been calculated in Ref.~\cite{Petraki:2016cnz}. It is found that in order for bound state formation to remain efficient, the binding energy of dark hydrogen $B_D$ must also be in the MeV range, which is the case for $\alpha_D \gtrsim 0.1$.

We find that for $B_D > m_\mathbf{A}$ recombination is in fact so efficient that the fraction of ionized particles is negligible at late times. For $B_D < m_\mathbf{A}$ on the other hand, the recombination process $\mathbf{p} + \mathbf{e} \to \mathbf{H} + \mathbf{A}$ is kinematically allowed only if the particles in the initial state have sufficient kinetic energy. Hence, this process becomes Boltzmann suppressed for temperatures $T < m_\mathbf{A} - B_D$ and the fraction of millicharged DM at late times can vary over many orders of magnitude.

At the same time, the dark photon can couple to SM particles via kinetic mixing with the visible photon. Because of this mixing the dark photon mediates interactions between DM particles and nuclei, which give rise to interesting experimental signatures. We calculate the form factor for the scattering of dark hydrogen off ordinary nuclei and show that direct detection experiments place a bound on the mass difference $m_\mathbf{p} - m_\mathbf{e}$ (see also Ref.~\cite{Cline:2012is}). Combining all of these considerations we identify the parameter regions that are consistent with existing constraints but predict exciting signals in future radio telescopes and direct detection experiments.

This letter is structured as follows. In Sec.~\ref{sec:model} we introduce the model that we consider and discuss how the dark photon mass and the DM millicharges arise simultaneously from the Stueckelberg mechanism. Sec.~\ref{sec:recombination} then provides details on the bound state formation and the evolution of the ionisation fraction of the dark sector. Potential signals in direct detection experiments and the resulting constraints are considered in Sec.~\ref{sec:DD}. We present our conclusions and discuss future directions in Sec.~\ref{sec:conclusions}.

\section{Model set-up}
\label{sec:model}

Our starting point is the usual set-up for atomic DM, i.e.\ we consider two different Dirac fermions $\mathbf{p}$ and $\mathbf{e}$ that carry opposite unit charge under a new $U(1)'$ gauge group with gauge boson $\mathbf{A}$:
\begin{equation}
 \mathcal{L} \supset - \mathbf{e} \mathbf{A}_\mu \left( \bar{\mathbf{p}} \gamma^\mu \mathbf{p} - \bar{\mathbf{e}} \gamma^\mu \mathbf{e} \right) \; .
\end{equation}
By definition, the dark proton is taken to be heavier than the dark electron ($m_\mathbf{p} > m_\mathbf{e}$), but we assume that the decay $\mathbf{p} \to \overline{\mathbf{e}} + \mathbf{A}^{(\ast)}$ is forbidden by some symmetry and hence both particles are stable. In analogy to the SM we allow for particle-antiparticle asymmetries~\cite{Kaplan:2009ag,Zurek:2013wia,Petraki:2013wwa} in the two components:
\begin{equation}
 \eta_{\mathbf{p},\mathbf{e}} \equiv \frac{n_{\mathbf{p},\mathbf{e}} - n_{\mathbf{\bar{p}},\mathbf{\bar{e}}}}{s} \neq 0 \; ,
\end{equation}
where $n$ and $s$ denote number and entropy density, respectively. In order to ensure overall charge neutrality, the two asymmetries must be equal: $\eta_\mathbf{p} = \eta_\mathbf{e}$. For $T \lesssim m_{\mathbf{p},\mathbf{e}} / 30$, the symmetric component will efficiently annihilate away via processes like $\mathbf{p} + \mathbf{\bar{p}} \to \mathbf{A} + \mathbf{A}$, and the relic abundance will be set by the initial asymmetries. We assume that $\Omega_\mathbf{p} + \Omega_\mathbf{e} \approx \Omega_\text{DM} = 0.12 / h^2$~\cite{Aghanim:2018eyx}, up to a small correction due to the energy release during recombination.

In contrast to the most commonly studied scenario, we consider the case that the dark photon $\mathbf{A}$ is massive. Specifically, we assume the presence of a Stueckelberg field $\sigma$, which under the gauge transformation $\mathbf{A}_\mu \to \mathbf{A}_\mu + \partial_\mu \phi(x)$ transforms as $\sigma \to \sigma + m_\mathbf{A} \phi(x)$. One can then write down the gauge-invariant Stueckelberg Lagrangian
\begin{equation}
 \mathcal{L}_\text{St} = \frac{1}{2} m_\mathbf{A}^2 (\mathbf{A}_\mu - \frac{1}{m_\mathbf{A}} \partial_\mu \sigma) (\mathbf{A}^\mu - \frac{1}{m_\mathbf{A}} \partial^\mu \sigma) \; ,
\end{equation}
which, together with a gauge fixing term, generates a mass term for the dark photon~\cite{Stueckelberg:1900zz}. Note that in the Stueckelberg mechanism the gauge symmetry remains unbroken and hence our assumption dark charge neutrality must be satisfied (see Ref.~\cite{Petraki:2014uza} for details).

It is well known that the exact same mechanism can be used to also generate millicharges for the dark fermions. To do so, we assume that the Stueckelberg field transforms under a gauge transformation of the electromagnetic gauge field $A^\mu \to A^\mu + \partial^\mu \phi(x)$ as $\sigma \to \sigma + \lambda m_\mathbf{A} \phi(x)$, where $\lambda$ is a free parameter. In this case the Stueckelberg field gives mass to a linear combination of the dark photon and the visible photon, while the orthogonal combination remains massless. For $\lambda \ll 1$ the massless gauge boson behaves almost exactly like the SM photon, except that it now couples to $\mathbf{p}$ and $\mathbf{e}$~\cite{Kors:2004dx,Feldman:2007wj}:
\begin{equation}
 \mathcal{L} \supset - \lambda \mathbf{e} A_\mu \left( \bar{\mathbf{p}} \gamma^\mu \mathbf{p} - \bar{\mathbf{e}} \gamma^\mu \mathbf{e} \right) \; .
\end{equation}
It is convenient to define $\epsilon \equiv \lambda \mathbf{e} / e$, where $e$ is the electromagnetic charge, such that the DM millicharge is given by $\epsilon e$.

We note that this Stueckelberg mixing is fundamentally different from the kinetic mixing $\tfrac{\kappa}{2} \mathbf{F}^{\mu\nu} F_{\mu\nu}$ with $\mathbf{F}^{\mu\nu} = \partial^\mu \mathbf{A}^\nu - \partial^\nu \mathbf{A}^\mu$, which does not give rise to millicharges. Nevertheless, kinetic mixing will in general also be present and modify the couplings of the dark photon to SM particles. A detailed discussion of this mixing is provided in~\ref{app:mixing}, where it is shown that for $\kappa \ll 1$ the dark photon couplings to SM particles are proportional to $\delta \equiv \lambda - \kappa$. The presence of this mixing is essential in order to ensure that the dark photons decay and do not overclose the Universe.

To summarize, our model contains two massive dark fermions with opposite charge and equal asymmetry, that couple to a massive dark photon as well as (very weakly) to the visible photon. The dark photon also couples very weakly to electrically charged SM fermions. The free parameters are hence the three masses $m_\mathbf{p}$, $m_\mathbf{e}$ and $m_\mathbf{A}$, the dark fine structure constant $\alpha_D = \mathbf{e}^2 / (4\pi)$, the millicharge $\epsilon$ and the effective mixing parameter $\delta$.

The millicharge $\epsilon$ is strongly constrained by the effect of DM-baryon scattering on the CMB~\cite{Dvorkin:2013cea,Wilkinson:2013kia,Slatyer:2018aqg,Boddy:2018wzy}. If all dark protons and electrons were to remain in the ionized state, these constraints would imply $\epsilon < 10^{-8}$, which would make it impossible to have observable effects from DM-baryon interactions at later times~\cite{Boddy:2018wzy}. However, dark photon exchange creates a Yukawa potential between dark protons and dark electrons, which allows for the formation of electrically neutral bound states (dark hydrogen $\mathbf{H}$) such that only a subcomponent of millicharged DM remains. The binding energies of these bound states can be calculated by numerically solving the Schr{\"o}dinger equation. For the 1s ground state one finds approximately~\cite{Petraki:2016cnz}
\begin{equation}
B_D \equiv m_\mathbf{p} + m_\mathbf{e} - m_\mathbf{H} \approx \left(1 - 0.84 \frac{m_\mathbf{A}}{\mu \alpha_D}\right)^{2.226} \frac{\mu \alpha_D^2}{2} \; ,
\label{eq:BD}
\end{equation}
where $\mu = m_\mathbf{p} m_\mathbf{e}  / (m_\mathbf{p} + m_\mathbf{e})$ is the reduced mass. 
%There are also a finite number of additional bound states with higher energy, but these do not play a role in the present context and will not be discussed further. 

As we will show in the following section, depending on the value of $B_D$ the late time ionisation fraction $f_\mathbf{e}$ can vary over many orders of magnitude. Previous studies of the $21\,\mathrm{cm}$ absorption profile have shown that the most interesting parameter region corresponds to $f_\mathbf{e} \sim 0.1\%$ and $10^{-5} \lesssim \epsilon \lesssim 10^{-4}$~\cite{Munoz:2018pzp,Kovetz:2018zan}. In this case we expect also the effective dark photon coupling $\delta$ to be greater than $10^{-5}$ (unless there is an accidental cancellation between $\lambda$ and $\kappa$). Experimental bounds from dark photon searches then require $m_\mathbf{A} \gtrsim 10\,\mathrm{MeV}$~\cite{Bjorken:2009mm,Bauer:2018onh}. At the same time, the masses of the millicharged DM particles should be as small as possible in order to enhance the effect of DM-baryon interactions. The main objective of the remainder of this letter will therefore be to identify the allowed mass ranges for $m_\mathbf{e}$ and $m_\mathbf{p}$.

\section{Bound state formation}
\label{sec:recombination}

In this section we calculate how efficiently bound states of dark protons and dark electrons form in the early Universe and what fraction of DM remains in the ionized state. We focus on the process $\mathbf{e}+\mathbf{p}\rightarrow \mathbf{H}+\mathbf{A}$ where $\mathbf{A}$ is an on-shell dark photon (alternative processes that may contribute to bound-state formation will be discussed below). For $B_D > m_\mathbf{A}$ the energy release from bound-state formation is large enough to create a dark photon even if the relative velocity $v_\text{rel}$ of the initial states vanish, whereas for $B_D < m_\mathbf{A}$ non-zero kinetic energy is required for the process to be allowed.

\begin{figure*}[t]
\centering
\includegraphics[width=1\columnwidth]{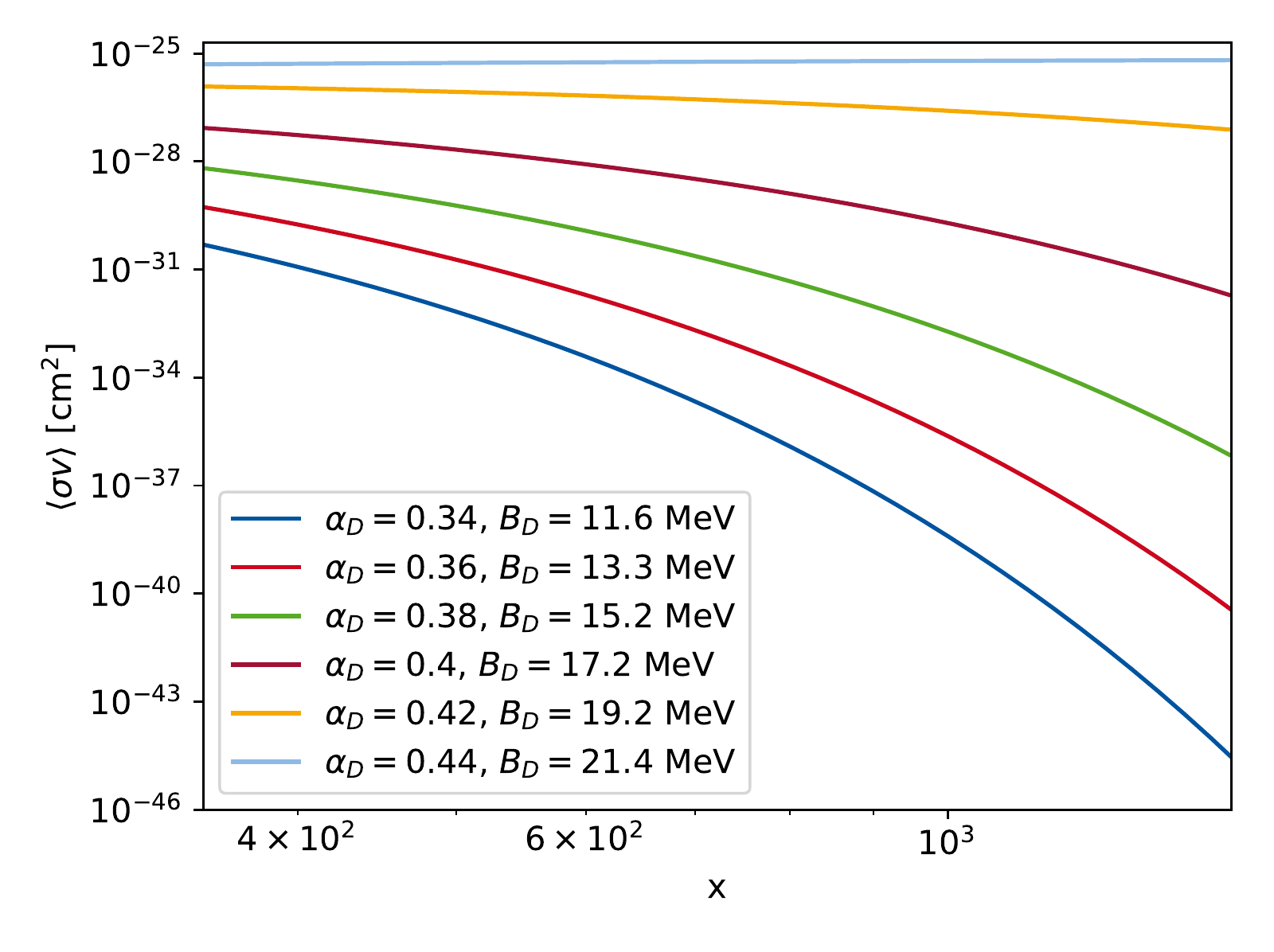}\qquad
\includegraphics[width=1\columnwidth]{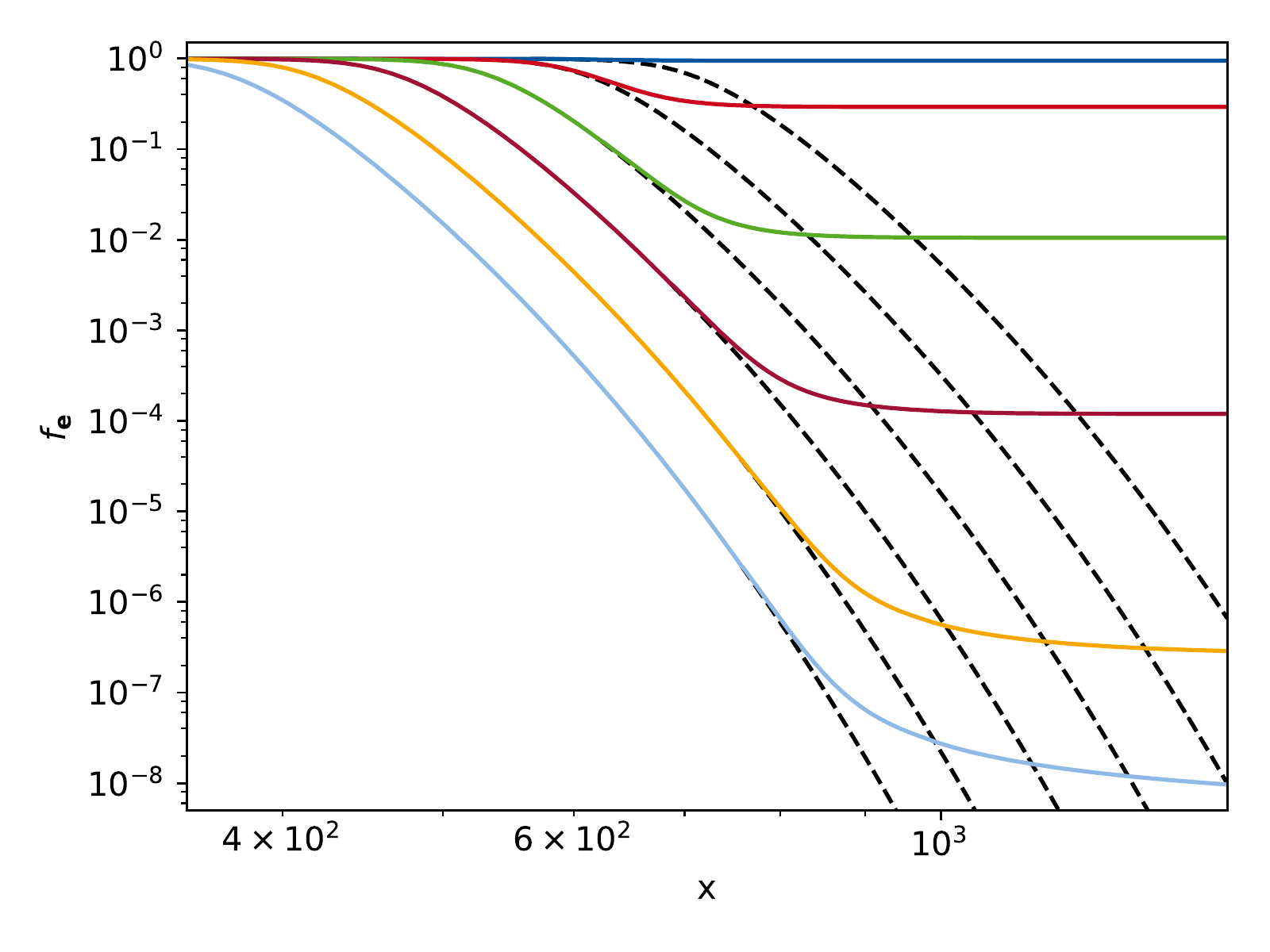}
\caption{\label{fig:Freeze_out} Evolution of the thermally averaged cross section for bound state formation (left) and the ionization fraction (right) as a function of the inverse temperature $x$ for $m_\mathbf{p}=750\,\textrm{MeV}$, $m_\mathbf{e}=500\,\textrm{MeV}$ and $m_\mathbf{A} = 20\,\mathrm{MeV}$ and different values of $\alpha_D$. The dashed lines in the right panel indicate the ionisation fraction in equilibrium. For binding energies smaller than the dark photon mass, bound state formation is suppressed at small temperatures and the present-day ionisation fraction increases.}
\end{figure*}

The cross section for bound state formation directly to the ground state is~\cite{Petraki:2016cnz}:
\begin{equation}
\label{eq:sigmav}
    \sigma v_{\textrm{rel}}=\frac{\pi \alpha_D^2}{8\mu^2}\sqrt{pss}\left(3-pss\right)\times S(v_{\textrm{rel}},\alpha_D, \mu, m_\mathbf{A}) \; ,
\end{equation}
where
\begin{equation}
 pss = 1 - \frac{4 m_\mathbf{A}^2}{\mu^2 v_\text{rel}^4 \left(1 + \frac{2 B_D}{\mu v_\text{rel}^2}\right)^2}
\end{equation}
is the phase space suppression factor, which vanishes for $\tfrac{1}{2} \mu v_\text{rel}^2 \leq m_\mathbf{A} - B_D$. The Sommerfeld factor $S(v_{\textrm{rel}},\alpha_D, \mu, m_\mathbf{A})$ must be calculated numerically and is taken from Ref.~\cite{Petraki:2016cnz}. Here and in the following we assume $\alpha_D\leq 0.8$ in order to ensure that the perturbative methods in Ref.~\cite{Petraki:2016cnz} can be applied. Of course, even for $\alpha_D$ slightly below this value perturbative methods already cease to be reliable and therefore our results in this parameter region should be interpreted with caution.

To calculate the rate of bound-state formation in the early Universe, we need to average the cross section from eq.~(\ref{eq:sigmav}) over a thermal distribution~\cite{Gondolo:1990dk,Cannoni:2013bza}: 
\begin{equation}
\langle \sigma v_\text{rel} \rangle=\frac{\int_{s_\text{min}}^\infty \textrm{d}s\, \sigma v_\text{rel} \beta \left(x^2/s\right)^{3/4} e^{-\sqrt{s}x/m}}{\sqrt{32\pi}(m m_\mathbf{e} m_\mathbf{p})^{3/2} e^{-(m_\mathbf{e}+m_\mathbf{p})x/m}} \; ,
\end{equation}
where $m\equiv m_\mathbf{p} \, m_\mathbf{e} / m_\mathbf{H}$, $x\equiv m/T$ with $T$ being the temperature of the thermal bath and $\sqrt{s}$ denotes the center-of-mass energy, which must be greater than $\sqrt{s_\text{min}} \equiv m_\mathbf{p}+m_\mathbf{e}+m_\mathbf{A}-B_D$ in order for $pss$ to be non-zero. Furthermore $\beta \equiv \left(s-m_\mathbf{e}^2-m_\mathbf{p}^2\right)\sqrt{\left[s-(m_\mathbf{e}+m_\mathbf{p})^2\right]\left[s-(m_\mathbf{e}-m_\mathbf{p})^2\right]}$ is introduced to simplify notation.

For $B_D > m_\mathbf{A}$ the thermally averaged cross section grows as the temperature decreases as a result of the growing Sommerfeld factor $S$. For $B_D < m_\mathbf{A}$ on the other hand, bound-state formation becomes exponentially suppressed as soon as the typical kinetic energy of particles in the thermal bath is insufficient to produce a dark photon, which is the case for $T \lesssim m_\mathbf{A} - B_D$.  This is illustrated in the left panel of Fig.~\ref{fig:Freeze_out}, which shows $\langle \sigma v \rangle$ as a function of $x$ for different values of $B_D$. 

Having calculated the bound-state formation cross section, we can obtain the dark ionization fraction\footnote{We use $n_\mathbf{e}$ to denote the number density of \emph{all} dark electrons, including  those bound in dark hydrogen, such that $n_\mathbf{H} \leq n_\mathbf{e}$.} $f_\mathbf{e}\equiv 1-n_\mathbf{H}/n_\mathbf{e}$ by integrating the Boltzmann equation
\begin{align}
    \frac{\textrm{d}f_\mathbf{e}}{\textrm{d}x}= - & \frac{\left<\sigma v\right>}{H} \left(\frac{1}{x}-\frac{1}{3g_{*s}}\frac{\textrm{d}g_{*s}}{\textrm{d}x}\right) \nonumber \\
    &\times \left[n_\mathbf{e} f_\mathbf{e}^2-\left(\frac{m^2}{2 \pi x}\right)^{3/2} e^{-B_D x/m}(1-f_\mathbf{e})\right],\label{Eq:Boltzmann}
\end{align}
where $H$ is the Hubble rate and $g_{\ast s}$ denotes the number of entropy degrees of freedom.\footnote{Note that the temperature dependence of $g_{\ast s}$ cannot be neglected, as $\frac{\textrm{d}g_{\ast s}}{\textrm{d}x}$ gives important contributions, especially at the time of electron-positron annihilation.} This equation is derived in \ref{app:Boltzmann} using the principle of detailed balance with the Saha equation serving as a boundary condition at high temperature. 

The resulting evolution of the ionization fraction as a function of inverse temperature is shown in the right panel of Fig.~\ref{fig:Freeze_out} for the same parameter combinations as in the left panel. In the case where $m_\mathbf{A} < B_D$, bound state formation is highly efficient and the fraction of dark electrons and dark protons that remain in the ionized state is suppressed to unobservable values. With decreasing $B_D$, however, bound state formation becomes less efficient and the present-day ionisation fraction can be sizable.

\begin{figure*}[t]
\centering
\includegraphics[width=0.99\columnwidth]{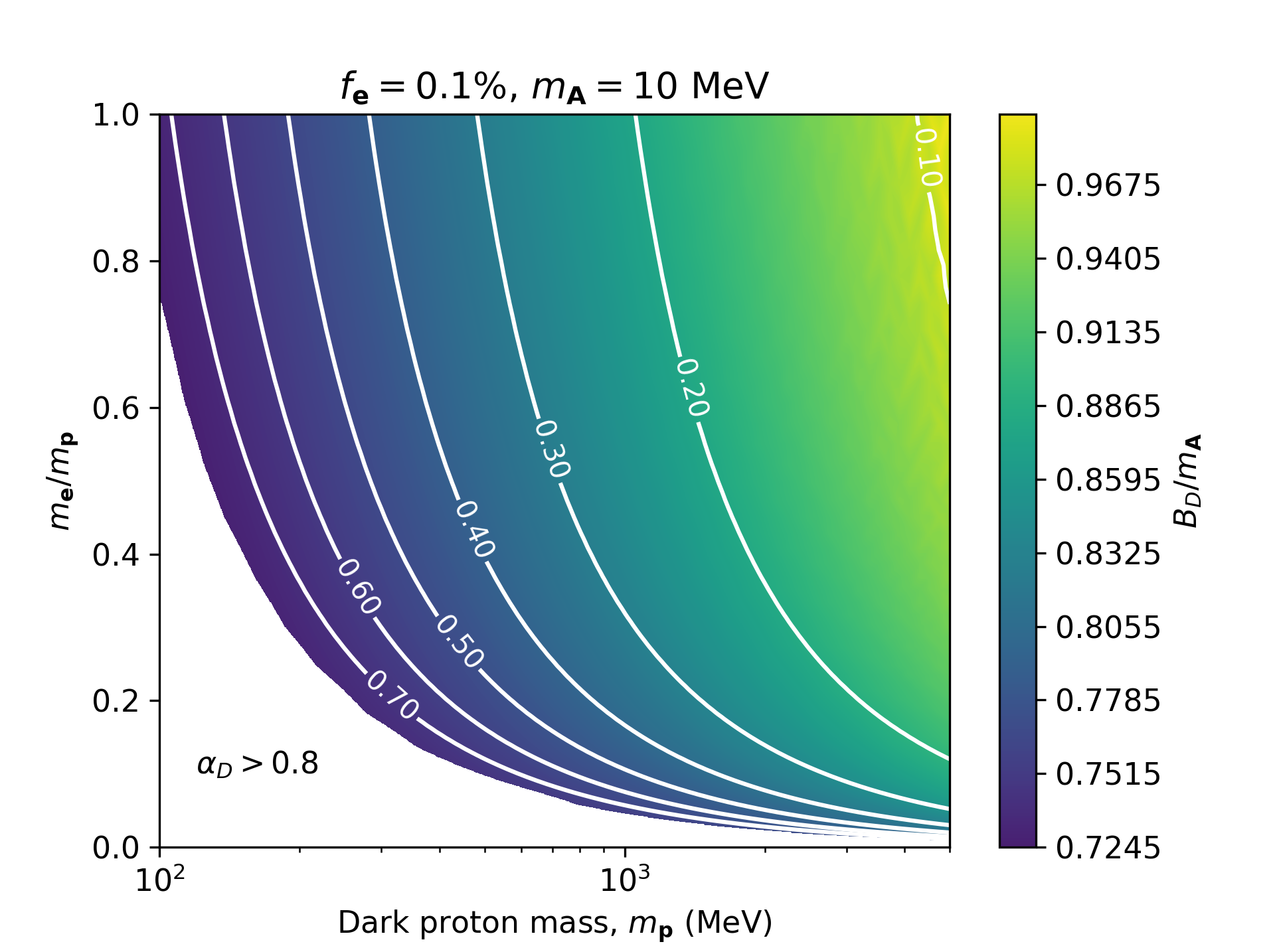}\qquad
\includegraphics[width=0.99\columnwidth]{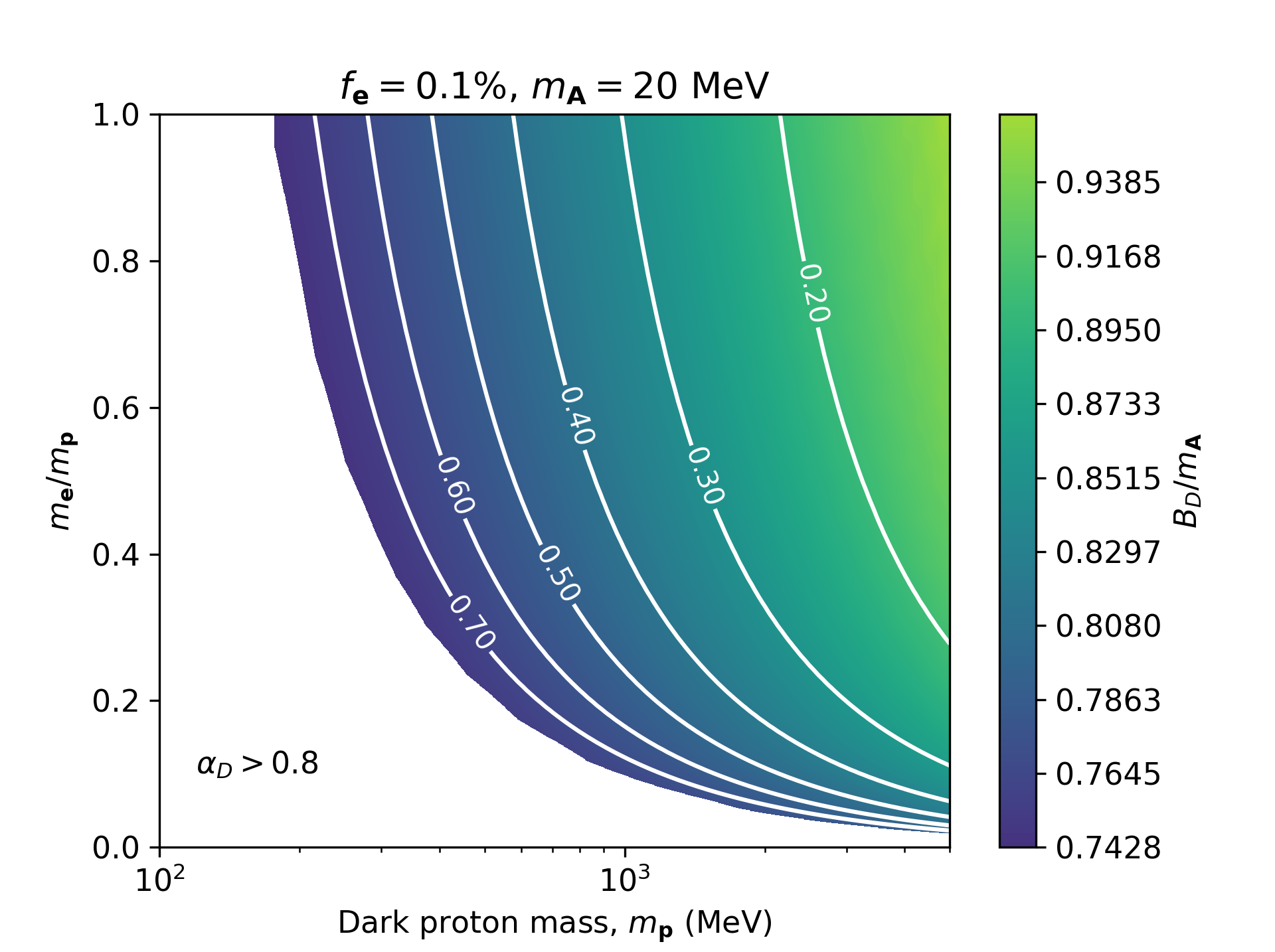}
\includegraphics[width=0.99\columnwidth]{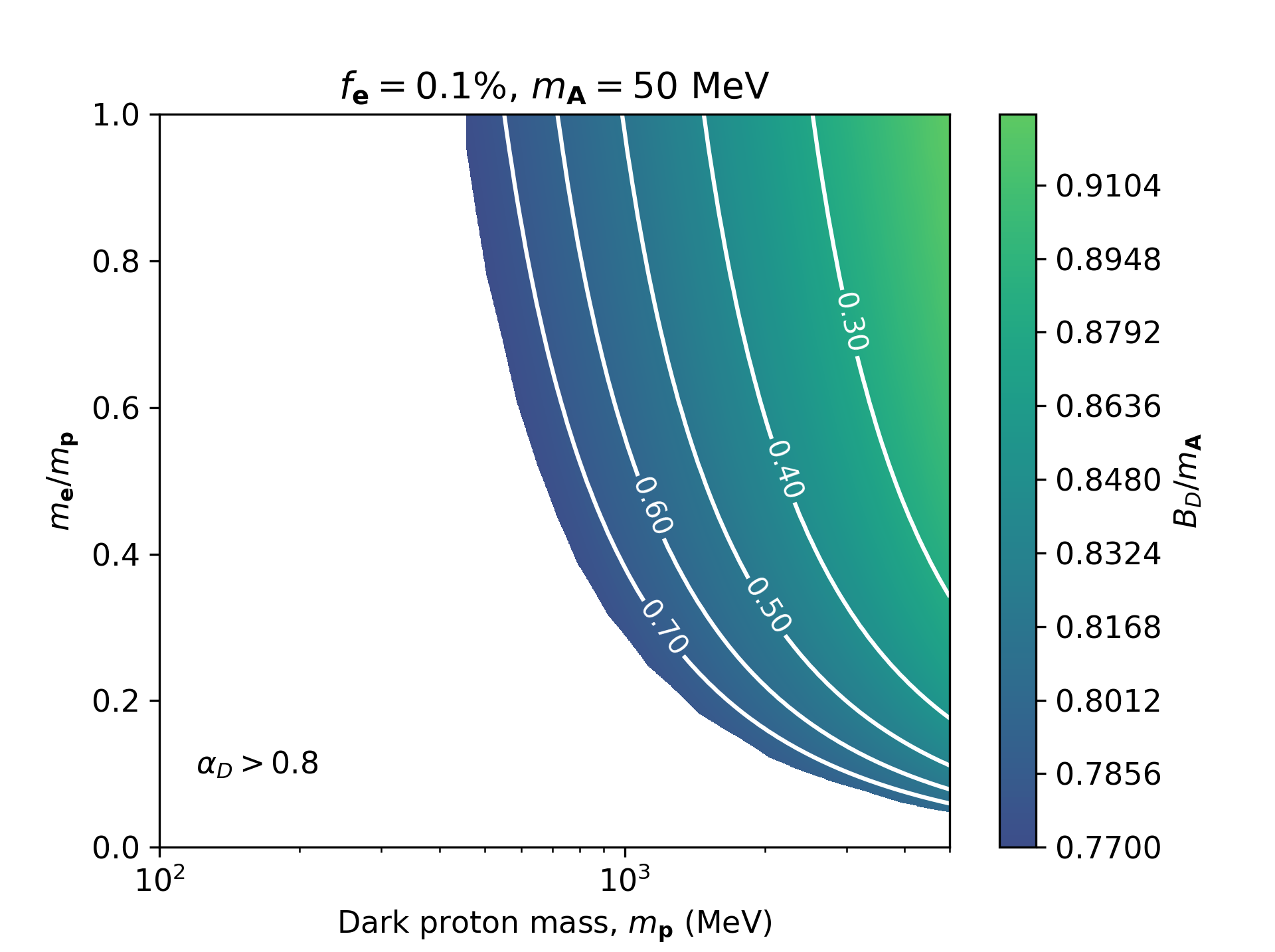}\qquad
\includegraphics[width=0.99\columnwidth]{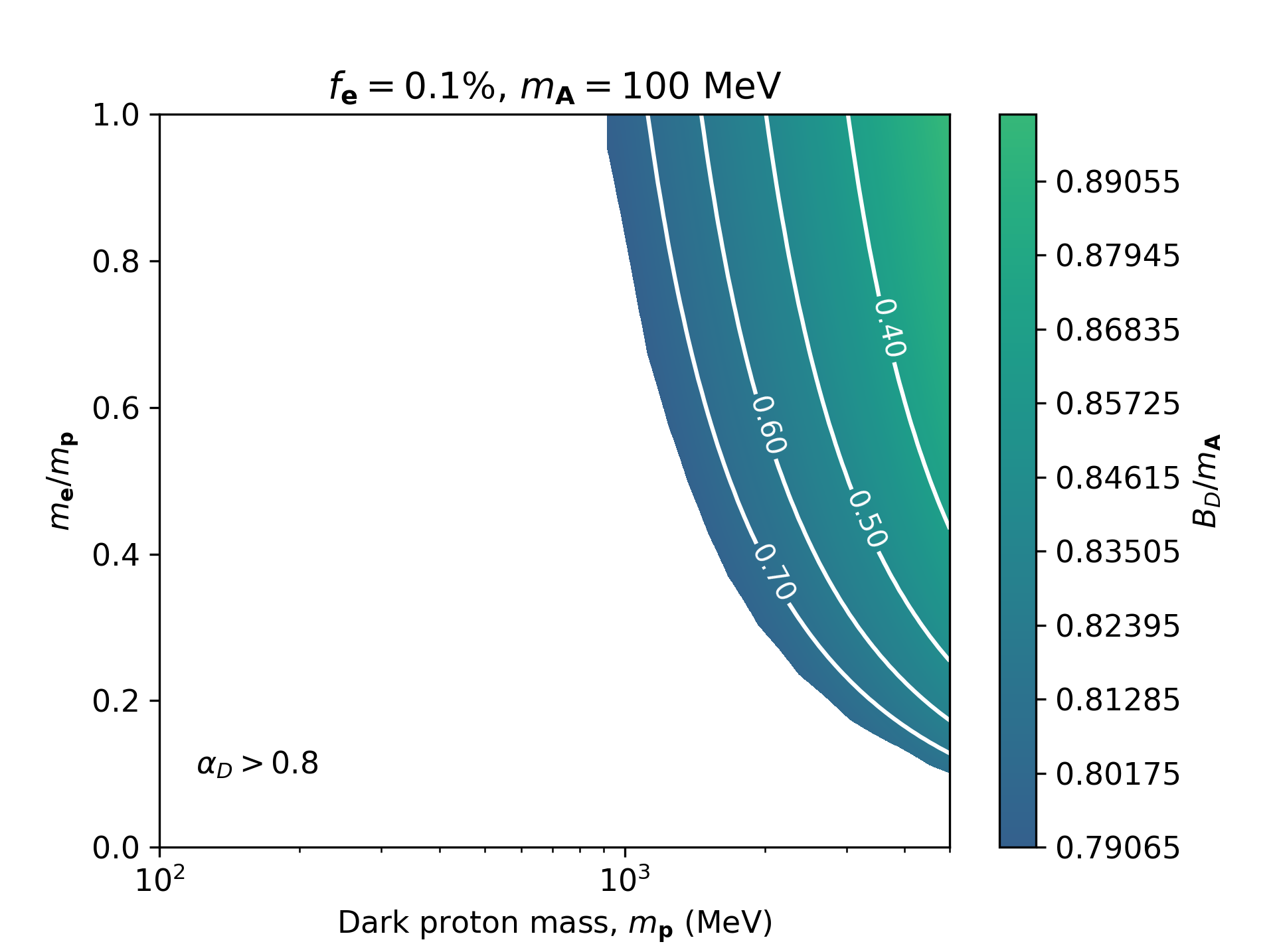}
\caption{\label{fig:B_D} Parameter combinations leading to a late-time ionisation fraction of $f_\mathbf{e} = 0.1\%$ for different values of the dark photon mass $m_\mathbf{A}$. The color shading represents the required binding energy $B_D$ in units of the dark photon mass, while the white contours indicate the corresponding values of $\alpha_D$. In the bottom-left corner in each panel it is impossible to achieve an ionisation fraction of $f_\mathbf{e} = 0.1\%$ with perturbative couplings ($\alpha_D < 0.8$).}
\end{figure*}

It should be clear from Fig.~\ref{fig:Freeze_out} that by varying $\alpha_D$ (and hence $B_D$) one can obtain essentially any value of $f_\mathbf{e}$ in the present Universe. We can invert this relation and determine the value of $B_D$ needed to produce a given ionisation fraction. The result of this procedure is shown in Fig.~\ref{fig:B_D} for the case $f_\mathbf{e} = 0.1\%$, which is small enough to be consistent with CMB constraints~\cite{Boddy:2018wzy} but potentially observable with future cosmological probes. The different panels correspond to different values of the dark photon mass $m_\mathbf{A}$. In each panel the bottom-left corner is excluded, because the required binding energy would correspond to  a non-perturbative coupling ($\alpha_D > 0.8$).

We find that in order to achieve an ionisation fraction of 0.1\%, the binding energy must lie in the range $0.7 < B_D / m_\mathbf{A} < 1$. Hence, larger dark photon masses require larger values of $B_D$, which implies either larger values of $\alpha_D$ (for fixed $m_\mathbf{e}$ and $m_\mathbf{p}$) or larger values of $m_\mathbf{e}$ and $m_\mathbf{p}$ (for fixed $\alpha_D$). With increasing dark photon mass the viable regions of parameter space therefore get pushed to larger and larger values of $m_\mathbf{p}$ and $m_\mathbf{e}$. In particular, given the requirement $m_\mathbf{A} \gtrsim 10\,\mathrm{MeV}$, it is impossible in our set-up to have $m_\mathbf{e} \lesssim 10\,\mathrm{MeV}$, a requirement independently imposed by considerations of the number of relativistic degrees of freedom during BBN~\cite{Creque-Sarbinowski:2019mcm} and the requirement of efficient annihilation $\mathbf{e}+\bar{\mathbf{e}} \to \mathbf{A} + \mathbf{A}$.

To conclude this discussion, let us briefly consider alternative processes that may contribute to bound-state formation. The process $\mathbf{p} + \mathbf{e} \to \mathbf{H}^\ast + \mathbf{A}$, where $\mathbf{H}^\ast$ denotes an excited state of dark hydrogen, requires even more kinetic energy in the initial state and is therefore strongly suppressed at low temperatures. The process $\mathbf{p} + \mathbf{e} \to \mathbf{H} + A$ with a visible photon $A$ is not suppressed at low temperatures, but it is suppressed by a factor $\epsilon^2$, which is much smaller than the Boltzmann suppression for dark photon emission in the relevant temperature range. Likewise the process $\mathbf{p} + \mathbf{e} \to \mathbf{H} + e^+ + e^-$ via an off-shell dark photon is suppressed proportional to $\delta^2$ and hence negligible. The same is true for bound state formation via scattering, $\mathbf{p} + \mathbf{e} + e^\pm \to \mathbf{H} + e^\pm$, which was recently studied in Ref.~\cite{Binder:2019erp}.

\section{Direct detection}
\label{sec:DD}

We have seen in the previous section that the DM particles in our set-up cannot be arbitrarily light. Indeed, even for $m_\mathbf{A} = 10\,\mathrm{MeV}$ it is impossible to have $m_\mathbf{H} < 150\,\mathrm{MeV}$, while for $m_\mathbf{A} = 50\,\mathrm{MeV}$ one finds $m_\mathbf{H} > 850 \, \mathrm{MeV}$. In this mass range it is essential to consider constraints from direct detection experiments searching for the scattering of DM particles off nuclei in low-background detectors.

At first sight, direct detection experiments place strong constraints on the millicharge $\epsilon$ of the ionized component, for which scattering can proceed through the exchange of visible photons. Following Ref.~\cite{Kahlhoefer:2017ddj} one obtains $\epsilon e^2 f_\mathbf{e} \lesssim 10^{-10}$ for $m_\mathbf{p} > 1 \, \mathrm{GeV}$, which for $f_\mathbf{e} = 0.1\%$ translates to $\epsilon \lesssim 10^{-6}$. However, Refs.~\cite{Chuzhoy:2008zy,McDermott:2010pa} argue that supernova explosions would expel millicharged DM particles from the Galactic disk for a wide range of $\epsilon$, so that the ionized component does not induce observable signals in terrestrial detectors.

The dominant direct detection constraint therefore arises from the scattering of dark hydrogen. Although these particles have vanishing net charge, they can have dipole interactions with ordinary nuclei via the exchange of either a visible or a dark photon. The latter contribution is in fact irreducible, since beam-dump experiments place a \emph{lower} bound on the mixing parameter $\delta$~\cite{Bjorken:2009mm,Bauer:2018onh}. For example, for $m_\mathbf{A} = 20\,\mathrm{MeV}$ ($m_\mathbf{A} = 50\,\mathrm{MeV}$), these constraints require $\delta > 2.2 \times 10^{-4}$ ($\delta > 1.7 \times 10^{-5}$), while $m_\mathbf{A} = 10\,\mathrm{MeV}$ is marginally excluded by a recent reanalysis of E774~\cite{Bauer:2018onh}.

To calculate the direct detection cross section we consider elastic scattering of a visible proton off the potential produced by a dark hydrogen atom,
\begin{equation}
		V(\mathbf{r})=\int \textrm{d}^3\Tilde{r} \frac{\sqrt{\alpha_D\alpha}\delta \rho(\Tilde{r})e^{-m_\mathbf{A}\left|\mathbf{r}-\Tilde{\mathbf{r}}\right|}}{\left|\mathbf{r}-\Tilde{\mathbf{r}}\right|} \; , \label{eq:V}
\end{equation}
where $\rho(r)=\rho_\mathbf{p}(r)-\rho_\mathbf{e}(r)$ with $\rho_{\mathbf{p},\mathbf{e}}(r)=\left(\alpha_D m_{\mathbf{p},\mathbf{e}}\right)^{3}  e^{-2r\alpha_D m_{\mathbf{p},\mathbf{e}}}/\pi$ is the charge distribution (see appendix~\ref{app:rho}). In the Born approximation one then obtains
\begin{align}
    \frac{\textrm{d}\sigma}{\textrm{d}q^2} = \, & \frac{\pi\alpha\delta^2q^4}{64\alpha_D^3 v^2 (m_\mathbf{A}^2+q^2)^2} \nonumber \\ 
    & \times \left| \frac{3 q^2}{\alpha_D^2m_\mathbf{e}^4}-\frac{3 q^2}{\alpha_D^2m_\mathbf{p}^4}-\frac{8}{m_\mathbf{e}^2}+\frac{8}{m_\mathbf{p}^2} \right| ^2, \label{dq2massive}
    \end{align}
where $v$ is the velocity of the incoming DM particle, $q$ is the recoil momentum, and we have neglected terms of higher order in $q$.\footnote{The cross section for the exchange of a visible photon can be obtained from this expression by taking $m_\mathbf{A}\rightarrow 0$, $\delta\rightarrow \epsilon$ and $\alpha_D \rightarrow \alpha$ in the first line.} We note that this expression differs significantly from the standard cases of spin-independent or spin-dependent scattering. In particular, $\mathrm{d}\sigma/\mathrm{d} q^2$ vanishes in the limit $q \to 0$.

%This gives 
%\begin{align}
%    \frac{\textrm{d}\sigma}{\textrm{d}q^2} = \, & \frac{\pi(1-\cos\theta)\alpha^2\epsilon^2\mu_{\mathbf{H},p}^2}{32q^2} \nonumber \\ & \times \left|  \frac{3 q^2}{\alpha_D^4m_\mathbf{e}^4}-\frac{3 q^2}{\alpha_D^4m_\mathbf{p}^4}-\frac{8}{\alpha_D^2m_\mathbf{e}^2}+\frac{8}{\alpha_D^2m_\mathbf{p}^2}\right| ^2. \label{dq2massless}
%\end{align}
An interesting limit to consider is that of $m_\mathbf{p}\approx m_\mathbf{e}$, which can be done by introducing $\Delta\equiv m_\mathbf{p}-m_\mathbf{e}$. At leading order in $\Delta$ the cross section becomes
\begin{equation}
    \frac{\textrm{d}\sigma}{\textrm{d}q^2}=\frac{\pi\alpha\delta^2\Delta^2 q^4}{4\alpha_D^7 v^2 m_\mathbf{p}^{10}} \left| \frac{4\alpha_D^2 m_\mathbf{p}^2 -3q^2}{m_\mathbf{A}^2+q^2} \right| ^2.
\end{equation}
It follows that the cross section tends to zero as $\Delta \to 0$ in the Born approximation, which is a direct consequence of $\rho(r)$ vanishing in this limit~\cite{Cline:2012is}.

\begin{figure}[t]
\centering
\includegraphics[width=0.95\columnwidth,clip,trim=5 5 35 30]{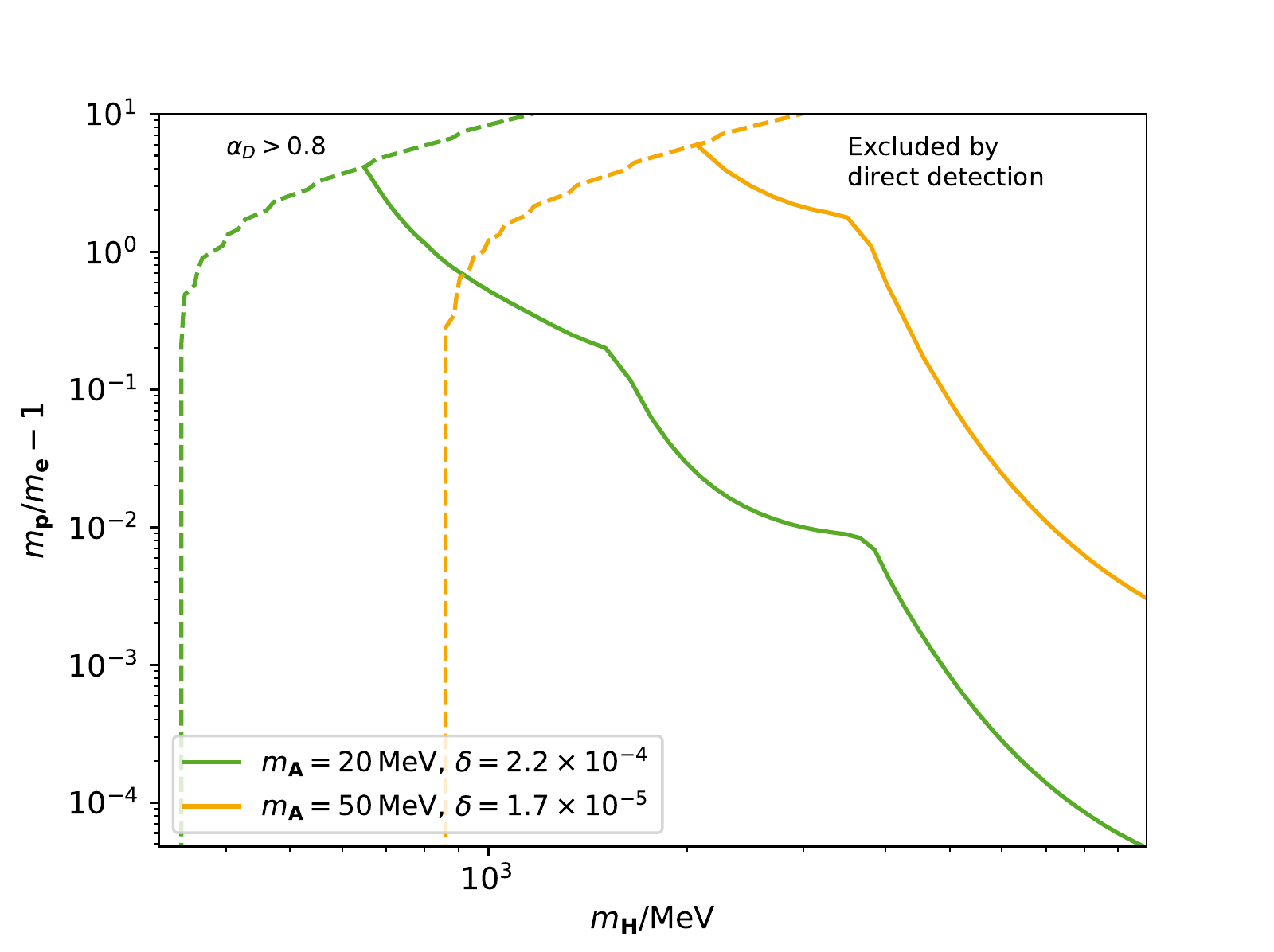}
\caption{\label{fig:Direct_detection} 
Bounds on the mass splitting $m_\mathbf{p} / m_\mathbf{e} - 1$ as a function of the dark hydrogen mass $m_\mathbf{H}$ for different values of the dark photon mass $m_\mathbf{A}$ and the mixing parameter $\delta$. Parameter regions above and to the right of the solid lines are excluded by direct detection experiments, while above and to the left of the dashed lines non-perturbative couplings ($\alpha_D > 0.8$) are necessary for efficient bound state formation. The plot assumes a present-day ionisation fraction of $f_\mathbf{e} = 0.1\%$ but the results only depend very slightly on this parameter.}
\end{figure}

For given values of $m_\mathbf{A}$, $m_\mathbf{H}$ and $\delta$ direct detection experiments therefore place an upper bound on the relative mass difference $\Delta / m_\mathbf{e} = m_\mathbf{p}/m_\mathbf{e} - 1$. To calculate these bounds, we have implemented the scattering cross section from eq.~(\ref{dq2massive}) in the public code \texttt{DDCalc\_v2}~\cite{Workgroup:2017lvb,Athron:2018hpc}, which then performs a combination of the exclusion limits from CRESST II~\cite{Angloher:2015ewa}, CDMSlite~\cite{Agnese:2015nto}, Xenon1T 2018~\cite{Aprile:2018dbl}, PandaX 2017~\cite{Cui:2017nnn} and PandaX 2016~\cite{Tan:2016zwf}. The results are shown in Fig.~\ref{fig:Direct_detection} for an ionisation fraction $f_\mathbf{e} = 0.1\%$ and for different values of $m_\mathbf{A}$. For each value of the dark photon mass, the mixing parameter has been set to the smallest value consistent with bounds from beam-dump experiments in order to give the most conservative bound. Stronger bounds would be obtained for larger values of $\delta$ as well as in the presence of an additional contribution from visible photon exchange.\footnote{Note that in principle there could be destructive interference between the contribution from dark photon exchange and the one from visible photon exchange. However, since the two contributions have different dependence on the momentum transfer $q$, it is impossible to significantly suppress the scattering rate.}

We also indicate in Fig.~\ref{fig:Direct_detection} the parameter regions where it is impossible to achieve $f_\mathbf{e} = 0.1\%$ with $\alpha_D < 0.8$. Together with the constraints from direct detection experiments, this requirement makes it impossible to have $m_\mathbf{p} \gg m_\mathbf{e}$. Indeed, as soon as the dark hydrogen mass is larger than a few GeV, $m_\mathbf{p}/m_\mathbf{e} - 1$ needs to be tuned to very small values to satisfy direct detection constraints. For smaller values of $m_\mathbf{H}$ direct detection constraints become weaker and additional parameter space opens up. For example, for $m_\mathbf{A} = 20\,\mathrm{MeV}$, one can have $m_\mathbf{p} = 500\,\mathrm{MeV}$, $m_\mathbf{e} = 200\,\mathrm{MeV}$ and $\alpha_D = 0.59$, leading to $m_\mathbf{H} = 685\,\mathrm{MeV}$. Interestingly, DM particles with these parameters would induce clear signals in future direct detection experiments like SuperCDMS~\cite{Agnese:2016cpb} and in accelerator experiments like FASER~\cite{Feng:2017uoz} or SHiP~\cite{Alekhin:2015byh}.

\section{Discussion and outlook}
\label{sec:conclusions}

In this work we have investigated the formation and detection bound states of dark protons $\mathbf{p}$ and dark electrons $\mathbf{e}$ that couple to a massive dark photon $\mathbf{A}$. The same mechanism that generates the dark photon mass also generates a millicharge for the dark protons and the dark electrons. At low temperatures most of the millicharged particles form neutral dark hydrogen $\mathbf{H}$ so that only a sub-dominant fraction of millicharged DM remains. This millicharged subcomponent gives rise to DM-baryon interactions that become stronger as the Universe cools down and, for certain values of the millicharge, the ionisation fraction and the DM mass, induce observable signals in the $21\,\mathrm{cm}$ absorption profile at cosmic dawn.

In order to determine the late-time ionisation fraction $f_\mathbf{e}$, we have calculated the thermally averaged cross section for bound state formation and solved the resulting Boltzmann equation. We find that $f_\mathbf{e}$ depends sensitively on the ratio of the binding energy of dark hydrogen and the dark photon mass and that it can vary over many orders of magnitude. By inverting this relation we identified the regions of parameter space where it is possible to obtain an ionisation fraction of $f_\mathbf{e} = 0.1\%$, which is interesting for $21\,\mathrm{cm}$ physics while being consistent with constraints from the CMB.

We have furthermore derived the differential cross section for the scattering of dark hydrogen off ordinary protons via dipole interactions and showed that this cross section is suppressed for small mass difference $\Delta = m_\mathbf{p} - m_\mathbf{e}$. Conversely, if the mass difference is sizable, constraints from direct detection experiments place a tight upper bound on the dark hydrogen mass. Combining this bound with the requirement of perturbative couplings then leads to a relatively small viable parameter range. For example, for dark photon masses close to current experimental bounds ($10\,\mathrm{MeV} \lesssim m_\mathbf{A} \lesssim  20\,\mathrm{MeV}$) this window is approximately given by $100\,\mathrm{MeV} \lesssim  m_\mathbf{H} \lesssim  1\,\mathrm{GeV}$ (see Fig.~\ref{fig:Direct_detection}).

With DM particles in this mass range it is not immediately possible to explain the EDGES anomaly. Nevertheless, it was pointed out recently~\cite{Liu:2019knx} that the effect of a millicharged subcomponent on the baryon temperature is enhanced if there is an additional force between the millicharged sub-component and the dominant neutral DM component. Such a force is automatically present in our set-up, because both components couple to dark photons. However, in contrast to what was assumed in Ref.~\cite{Liu:2019knx} the force mediated by dark photons does not become stronger at small velocities, because the dark hydrogen only has dipole interactions. It will be very interesting to investigate whether such a short-range force can also help to alleviate the tension between the EDGES anomaly and other cosmological observations.

Another interesting direction for future research is the effect of scattering between dark hydrogen and cosmic rays, which has been shown to be relevant for direct detection of sub-GeV DM particles~\cite{Bringmann:2018cvk}. In the present context, collisions with cosmic rays could potentially overcome the binding energy of dark hydrogen and produce a flux of energetic dark electrons and dark protons, which would leave striking signals in direct detection and neutrino experiments. Likewise, collisions with cosmic rays could induce excitations of dark hydrogen, with the subsequent de-excitation leading to potentially observable signatures in the $\gamma$-ray sky.

Finally, the new decade promises a wealth of new experiments targeted at the direct detection of sub-GeV DM and the search for dark photons. These searches are highly complementary in the sense that dark photon searches are most sensitive for small values of $m_\mathbf{A}$ while direct detection experiments give the strongest constraints for large $m_\mathbf{H}$. The parameter space of the model studied here will therefore soon be probed comprehensively and we can look forward to learning more about the interplay between particle DM and 21-cm cosmology.

\section*{Acknowledgements}

We are grateful to Eike M{\"u}ller for collaboration at the early stages of this project and to Riccardo Catena for offering advice on the calculation of the effective charge distribution. We also thank Tobias Binder, Michael Kachelriess and Julien Lesgourgues for discussions and Torsten Bringmann, Camilo Garcia-Cely, Fatih Ertas and Kai Schmidt-Hoberg for helpful comments on the manuscript. This work is funded by the DFG Emmy Noether Grant No.\ KA 4662/1-1.

\appendix

\section{Derivation of photon mixing}
\label{app:mixing}

In this appendix we derive the transformation of the interaction eigenstates $\tilde{A}$ and $\tilde{\mathbf{A}}$ needed to diagonalize the kinetic and mass terms~\cite{Eike}. Defining
\begin{equation}
 \tilde{\mathcal{A}}^\mu = 
 \begin{pmatrix} 
 \tilde{\mathbf{A}}^\mu \\ \tilde{A}^\mu
\end{pmatrix}
\end{equation}
and $\tilde{\mathcal{F}}_{\mu\nu} = \partial_\mu \tilde{\mathcal{A}}_\nu - \partial_\nu \tilde{\mathcal{A}}_\mu$, we can write the kinetic and mass terms before diagonalization as
\begin{equation}
 \mathcal{L} = - \frac{1}{4} \tilde{\mathcal{F}}_{\mu\nu}^T \mathcal{K} \tilde{\mathcal{F}}^{\mu\nu} - \frac{1}{2} \tilde{\mathcal{A}}_\mu^T \mathcal{M}^2 \tilde{\mathcal{A}}^\mu
\end{equation}
with
\begin{equation}
 \mathcal{K} =
  \begin{pmatrix} 
 1 & \kappa \\ \kappa & 1
\end{pmatrix} \, , \qquad 
 \mathcal{M}^2 =
  \begin{pmatrix} 
 m_\mathbf{A}^2 & \lambda m_\mathbf{A}^2 \\ \lambda m_\mathbf{A}^2 & \lambda^2 m_\mathbf{A}^2
\end{pmatrix} \; .
\end{equation}
The kinetic term is diagonalized by the general linear transformation\footnote{Note that in order to simultaneously diagonalise both mass term and kinetic term, $\mathcal{G}$ must be a general linear transformation (rather than an orthogonal one), which explains why kinetic mixing alone does not lead to a DM millicharge.} $\tilde{\mathcal{A}} \to \mathcal{G} \tilde{\mathcal{A}}$ with
\begin{equation}
 \mathcal{G} =
  \begin{pmatrix} 
 1 & - \frac{\kappa}{\sqrt{1-\kappa^2}} \\ 0 & \frac{1}{\sqrt{1-\kappa}}
\end{pmatrix} \; ,
\end{equation}
which leads to the modified mass matrix
\begin{equation}
 \mathcal{G}^T \mathcal{M}^2 \mathcal{G} = 
  \begin{pmatrix} 
 m_\mathbf{A}^2 & \frac{\lambda - \kappa}{\sqrt{1-\kappa^2}} m_\mathbf{A}^2 \\ \frac{\lambda - \kappa}{\sqrt{1-\kappa^2}} m_\mathbf{A}^2 & \frac{(\lambda - \kappa)^2}{1-\kappa^2} m_\mathbf{A}^2 
\end{pmatrix} \; .
\end{equation}
To diagonalize the mass matrix we require an orthogonal transformation $\mathcal{G} \tilde{\mathcal{A}} \to \mathcal{O} \mathcal{G} \tilde{\mathcal{A}}$ with
\begin{equation}
 \mathcal{O} = 
  \begin{pmatrix} 
 \cos \theta & - \sin \theta \\ \sin \theta & \cos \theta
\end{pmatrix}
\end{equation}
and $\tan \theta = (\lambda - \kappa) / \sqrt{1 - \kappa^2}$.  The mass eigenstates are then given by $\mathcal{A} = \mathcal{O} \mathcal{G} \tilde{\mathcal{A}}$, and the mass eigenvalues are found to be $m_\mathbf{A}^2 (1 - 2 \kappa \lambda + \lambda^2) / (1 - \kappa^2) \approx m_\mathbf{A}^2$ and 0. We then find at leading order in $\kappa$ and $\lambda$:
\begin{equation}
 \tilde{\mathcal{A}} = 
   \begin{pmatrix} 
 1 & \lambda \\ \kappa - \lambda & 1
\end{pmatrix}
\mathcal{A} \; .
\end{equation}
From this mixing matrix we can directly read off that the visible photon obtains a coupling $\lambda \mathbf{e}$ to DM particles, while the dark photon obtains a coupling $(\kappa - \lambda) q e$ to SM particles with charge $q$.

\section{Derivation of the Boltzmann Equation}
\label{app:Boltzmann}

In this appendix we derive eq.~(\ref{Eq:Boltzmann}), which governs the evolution of the ionization fraction. At high temperatures, the reaction $\mathbf{p}+\mathbf{e}\rightleftharpoons\mathbf{H}+\mathbf{A}$ is efficient enough to keep $\mathbf{H}$ in equilibrium, and the equilibrium ionization fraction $f^\text{eq}_\mathbf{e}$ is governed by the Saha equation
\begin{equation}
\label{eq:Saha}
1-f^\text{eq}_\mathbf{e} = n_\mathbf{e} (f_\mathbf{e}^\text{eq})^2 \left(\frac{2 \pi x}{m^2}\right)^{3/2} e^{B_D x/m} \; .
\end{equation}
Away from equilibrium, the time derivative of the ionisation fraction $\Dot{f_\mathbf{e}}$ is determined by the rates of $\mathbf{e}+\mathbf{p}\rightarrow\mathbf{H}+\mathbf{A}$ and $\mathbf{H}+\mathbf{A}\rightarrow\mathbf{e}+\mathbf{p}$. The first process creates dark hydrogen at a rate of $\left<\sigma v\right>n_\mathbf{e}^2 f_\mathbf{e}^2$, while the second one destroys dark hydrogen at a rate proportional to $(1 - f_\mathbf{e}) n_\mathbf{e}$. Hence we can write
\begin{equation}
    -n_\mathbf{e}\Dot{f_\mathbf{e}}=\left<\sigma v\right>n_\mathbf{e}^2 f_\mathbf{e}^2- n_\mathbf{e} (1 - f_\mathbf{e}) \zeta(x) \; .\label{Eq:detailed_balance}
\end{equation}
with some function $\zeta(x)$. In equilibrium $\Dot{f_\mathbf{e}}=0$ and the Saha equation is satisfied. By comparing eq.~\eqref{Eq:detailed_balance} and eq.~\eqref{eq:Saha}, the function $\zeta$(x) can hence be identified, resulting in the Boltzmann equation:
\begin{equation}
    -\Dot{f_\mathbf{e}}=\left<\sigma v\right>\left[n_\mathbf{e} f_\mathbf{e}^2-\left(\frac{m^2}{2 \pi x}\right)^{3/2} e^{-B_D x/m}(1-f_\mathbf{e})\right]
    \label{Eq:Boltzmann(f)}.
\end{equation}
It is convenient to replace the time derivative of the ionization fraction by a derivative with respect to $x$, which depends on time as
\begin{equation}
    x \propto g_{*s}(t)^{1/3} a(t) \; .
\end{equation}
This gives
\begin{equation}
    \frac{\Dot{x}}{x}=\frac{\Dot{a}}{a}+\frac{1}{3}\frac{\Dot{g}_{*s}}{g_{*s}}=H+\frac{\Dot{x}}{3g_{*s}}\frac{\textrm{d}g_{*s}}{\textrm{d}x}
\end{equation}
and thus
\begin{equation}
    \Dot{x}=\frac{H}{\frac{1}{x}-\frac{1}{3g_{*s}}\frac{\textrm{d}g_{*s}}{\textrm{d}x}} \; .\label{Eq:xdot}
\end{equation}
Inserting \eqref{Eq:xdot} in \eqref{Eq:Boltzmann(f)} gives
\begin{align}
    \frac{\textrm{d}f_\mathbf{e}}{\textrm{d}x}= - & \frac{\left<\sigma v\right>}{H} \left(\frac{1}{x}-\frac{1}{3g_{*s}}\frac{\textrm{d}g_{*s}}{\textrm{d}x}\right) \nonumber \\
    &\times \left[n_\mathbf{e} f_\mathbf{e}^2-\left(\frac{m^2}{2 \pi x}\right)^{3/2} e^{-B_D x/m}(1-f_\mathbf{e})\right].
\end{align}
Note that we have neglected the effect of dark anti-protons and anti-electrons, which will modify the discussion at temperatures above the freeze-out temperature ($x \sim 30$). However, the temperature range relevant for dark recombination is significantly lower (see Fig.~\ref{fig:Freeze_out}) and it is therefore justified to neglect this contribution.

\begin{widetext}

\section{Derivation of the effective charge distribution}
\label{app:rho}

In this appendix we derive the effective charge distribution $\rho(r)$ that enters into eq.~(\ref{eq:V}). We start from the general expression for the scattering amplitude in the Born approximation
\begin{equation}
 f(\mathbf{q}) = \frac{\mu_{\mathbf{H},p}}{2\pi} \int \mathrm{d}^3 x' e^{- i \mathbf{q} \cdot \mathbf{x}'} V(\mathbf{x}') \; ,
\end{equation}
where $\mu_{\mathbf{H},p}$ denotes the reduced mass of the dark hydrogen and the visible proton.
In our case the potential receives two contributions: One from dark proton scattering and one from dark electron scattering. Due to the opposite charges of dark proton and dark electron, these two contributions have opposite sign. Hence we can write
\begin{equation}
 V(\mathbf{x}') = \sqrt{\alpha \alpha_D} \delta \int \mathrm{d}^3 P' \mathrm{d}^3 x_{\mathbf{p}} \mathrm{d}^3 x_{\mathbf{e}} \left[ \frac{e^{- | \mathbf{x}_{\mathbf{p}} - \mathbf{x}' | m_{\mathbf{A}}}}{| \mathbf{x}_{\mathbf{p}} - \mathbf{x}' |} -  \frac{e^{- | \mathbf{x}_{\mathbf{e}} - \mathbf{x}' | m_{\mathbf{A}}}}{| \mathbf{x}_{\mathbf{e}} - \mathbf{x}' |} \right] \Psi^\ast_2(\mathbf{x}_{\mathbf{p}},  \mathbf{x}_{\mathbf{e}}) \Psi_1(\mathbf{x}_{\mathbf{p}},  \mathbf{x}_{\mathbf{e}}) \; ,
\end{equation}
where $\mathbf{x}_{\mathbf{p},\mathbf{e}}$ denote the position of dark electron and dark proton, $\mathbf{P}'$ is the centre-of-mass momentum after scattering and $\Psi_{1,2}(\mathbf{x}_{\mathbf{p}},  \mathbf{x}_{\mathbf{e}})$ denote the wave function of dark hydrogen before and after scattering. By appropriate transformations of $\mathbf{x}'$ we then obtain
\begin{equation}
 f(\mathbf{q}) = \frac{\mu_{\mathbf{H},p}}{2\pi} \sqrt{\alpha \alpha_D}\delta \int \mathrm{d}^3 x' e^{- i \mathbf{q} \cdot \mathbf{x}'}  \frac{e^{- r' m_{\mathbf{A}}}}{r'} 
 \int \mathrm{d}^3 P' \mathrm{d}^3 x_{\mathbf{p}} \mathrm{d}^3 x_{\mathbf{e}} \left[e^{i \mathbf{q} \cdot \mathbf{x}_{\mathbf{p}}} - e^{i \mathbf{q} \cdot \mathbf{x}_{\mathbf{e}}}\right] \Psi^\ast_2(\mathbf{x}_{\mathbf{p}},  \mathbf{x}_{\mathbf{e}}) \Psi_1(\mathbf{x}_{\mathbf{p}},  \mathbf{x}_{\mathbf{e}}) \; ,
\end{equation}
where $r' = |\mathbf{x}'|$. We note that the same expression can also be obtained by calculating the scattering amplitude for $3\to3$ scattering of a dark proton, a dark electron and a nucleus and then dressing the amplitude with the appropriate dark hydrogen wave functions.

Now we transform to the centre-of-mass coordinates $\mathbf{x}_\text{cm} = (m_{\mathbf{p}} \mathbf{x}_{\mathbf{p}} + m_{\mathbf{e}} \mathbf{x}_{\mathbf{e}})/(m_{\mathbf{p}} + m_{\mathbf{e}})$ and $\mathbf{x}_\text{rel} = \mathbf{x}_{\mathbf{p}} - \mathbf{x}_{\mathbf{e}}$, in which the dark hydrogen wave functions can be written as
\begin{align}
 \Psi_1(\mathbf{x}_\text{cm}, \mathbf{x}_\text{rel}) & = e^{i \mathbf{P} \cdot \mathbf{x}_\text{cm}} \psi(\mathbf{x}_\text{rel}) \; , \\
 \Psi_2(\mathbf{x}_\text{cm}, \mathbf{x}_\text{rel}) & = e^{i \mathbf{P}' \cdot \mathbf{x}_\text{cm}} \psi(\mathbf{x}_\text{rel})
\end{align}
with $\mathbf{P}$ being the centre-of-mass momentum before scattering and $\psi(\mathbf{x}_\text{rel})$ denoting the ground-state wave function of the dark hydrogen atom. These expressions lead to
\begin{equation}
 \Psi^\ast_2(\mathbf{x}_{\mathbf{p}},  \mathbf{x}_{\mathbf{e}}) \Psi_1(\mathbf{x}_{\mathbf{p}},  \mathbf{x}_{\mathbf{e}}) = e^{i (\mathbf{P} - \mathbf{P}') \cdot \mathbf{x}_\text{cm}} |\psi(\mathbf{x}_\text{rel})|^2 \; .
\end{equation}
Now we note that $\mathbf{x}_{\mathbf{e}} - \mathbf{x}_\text{cm} = - m_{\mathbf{p}} \mathbf{x}_\text{rel} / (m_{\mathbf{p}} + m_{\mathbf{e}}) = - \mu \mathbf{x}_\text{rel} / m_{\mathbf{e}}$ and $\mathbf{x}_{\mathbf{p}} - \mathbf{x}_\text{cm} = \mu \mathbf{x}_\text{rel} / m_{\mathbf{p}}$, so that we obtain
\begin{align}
 f(\mathbf{q}) = \frac{\mu_{\mathbf{H},p}}{2\pi} \sqrt{\alpha \alpha_D}\delta \int &  \mathrm{d}^3 x' e^{- i \mathbf{q} \cdot \mathbf{x}'}  \frac{e^{- r' m_{\mathbf{A}}}}{r'} \int \mathrm{d}^3 P' \mathrm{d}^3 x_\text{cm} e^{- i (\mathbf{q} + \mathbf{P} - \mathbf{P}') \cdot \mathbf{x}_\text{cm}}  \nonumber \\ & \times \int \mathrm{d}^3 x_\text{rel} \left[e^{i \tfrac{\mu}{m_{\mathbf{p}}} (\mathbf{P}' - \mathbf{P}) \cdot \mathbf{x}_\text{rel}} - e^{- i \tfrac{\mu}{m_{\mathbf{e}}} (\mathbf{P}' - \mathbf{P}) \cdot \mathbf{x}_\text{rel}} \right] |\psi(\mathbf{x}_\text{rel})|^2 \; .
\end{align}
The integral over $\mathrm{d}^3 x_\text{cm}$ then yields $\delta^{3}(\mathbf{q} + \mathbf{P} - \mathbf{P}')$, which enables us to perform the integration over $\mathrm{d}^3 P'$ and impose momentum conservation: $\mathbf{P}' - \mathbf{P} = \mathbf{q}$. We transform the integration over $\mathrm{d}^3 x_\text{rel}$ by defining $\mathbf{x} = \tfrac{\mu}{m_{\mathbf{p}}} \mathbf{x}_\text{rel}$ in the first and $\mathbf{x} = - \tfrac{\mu}{m_{\mathbf{e}}} \mathbf{x}_\text{rel}$ in the second part of the integral. This yields 
\begin{align}
 f(\mathbf{q}) = \frac{\mu_{\mathbf{H},p}}{2\pi} \sqrt{\alpha \alpha_D}\delta \int \mathrm{d}^3 x' e^{- i \mathbf{q} \mathbf{x}'}  \frac{e^{- r' m_{\mathbf{A}}}}{r'} \int \mathrm{d}^3 x e^{i \mathbf{q} \cdot \mathbf{x}} \left[ \frac{m_{\mathbf{p}}^3}{\mu^3} \left\lvert\psi\left(\frac{m_{\mathbf{p}}}{\mu} \mathbf{x} \right)\right\lvert^2 - \frac{m_{\mathbf{e}}^3}{\mu^3} \left\lvert\psi\left(\frac{m_{\mathbf{e}}}{\mu} \mathbf{x}  \right)\right\lvert^2 \right] \; .
\end{align}

Finally, we again shift the integration variable $\mathbf{x}'$ to obtain
\begin{align}
 f(\mathbf{q}) = & \frac{\mu_{\mathbf{H},p}}{2\pi} \sqrt{\alpha \alpha_D}\delta \int \mathrm{d}^3 x' e^{- i \mathbf{q} \mathbf{x}'} \int \mathrm{d}^3 x \frac{e^{- | \mathbf{x} - \mathbf{x}' | m_{\mathbf{A}}}}{| \mathbf{x} - \mathbf{x}' |} \rho(r)
 \label{eq:fofq} \end{align}
with
\begin{equation}
\rho(r) = \frac{m_{\mathbf{p}}^3}{\mu^3} \left\lvert\psi\left(\frac{m_{\mathbf{p}}}{\mu} \mathbf{x} \right)\right\lvert^2 - \frac{m_{\mathbf{e}}^3}{\mu^3} \left\lvert\psi\left(\frac{m_{\mathbf{e}}}{\mu} \mathbf{x}  \right)\right\lvert^2  \; .
\end{equation}
Strictly speaking the wave function $\psi(r)$ needs to be calculated by solving the Schroedinger equation for a Yukawa potential. However, since the Bohr radius $\alpha_D \mu$ is significantly larger than the dark photon mass, we can approximate the Yukawa potential by a Coloumb potential and obtain
\begin{equation}
 \psi(r) = \frac{(\alpha_D \mu)^{3/2}}{\sqrt{\pi}} e^{-r \alpha_D \mu} \; ,
\end{equation}
which leads to
\begin{equation}
\rho(r) = \frac{\alpha_D^3 m_{\mathbf{p}}^3}{\pi} e^{-2 r \alpha_D m_{\mathbf{p}}} - \frac{\alpha_D^3 m_{\mathbf{e}}^3}{\pi} e^{-2 r \alpha_D m_{\mathbf{e}}}  \; . 
\end{equation}

Eq.~(\ref{eq:fofq}) can be rewritten as
\begin{equation}
 f(\mathbf{q}) = \frac{2 \delta \sqrt{\alpha \alpha_D} \mu_{\mathbf{H},p}}{\mathbf{q}^2 + m_{\mathbf{A}}^2} \tilde{\rho}(\mathbf{q}) \; , 
\end{equation}
with
\begin{align}
\tilde{\rho}(\mathbf{q}) & = \frac{4\pi}{q} \int_0^\infty \mathrm{d} r r \rho(r) \sin(q r) \nonumber \\
& = 16 \alpha_D^4 \left[ \frac{m_\mathbf{p}^4}{(q^2 + 4 m_\mathbf{p}^2 \alpha_D^2)^2} - \frac{m_\mathbf{e}^4}{(q^2 + 4 m_\mathbf{e}^2 \alpha_D^2)^2} \right] \nonumber \\
& \approx \frac{q^2}{16 \alpha_D^2} \left[ \frac{3 q^2}{\alpha_D^2 m_\mathbf{p}^4} - \frac{3 q^2}{\alpha_D^2 m_\mathbf{e}^4} + \frac{8}{m_\mathbf{e}^2} - \frac{8}{m_\mathbf{p}^2} \right]\; , 
\end{align}
where the approximation in the final step is valid for $q \ll \alpha_D m_{\mathbf{p},\mathbf{e}}$.

Eq.~(\ref{dq2massive}) now follows from
\begin{equation}
 \frac{\mathrm{d}\sigma}{\mathrm{d}q^2} = \frac{\pi}{k^2} |f(q)|^2
\end{equation}
with $k = \mu_{\mathbf{H},p} v$. 

\vspace{3mm}

\end{widetext}

\end{document}